%% file: astroph.tex
\documentclass{article}
\usepackage{emulateapj}
\usepackage{apjfonts}
\usepackage{graphicx}
\usepackage{onecolfloat}
\submitted{accepted for publication in ApJ}
\newcommand{\csstar}{CS~31062}
\newcommand{\gfvalue}{$gf$-value}
\newcommand{\gfvalues}{$gf$-values}
\newcommand{\Sec}{${}^{\prime\prime}$\llap{.}}
\newcommand{\etal}{et al.}
\newcommand{\gtsim}{$\scriptscriptstyle \; \buildrel > \over \sim \;$} 
\newcommand{\thch}{$^{13}$CH}
\newcommand{\sss}{$s_{ss}$}
\newcommand{\rss}{$r_{ss}$}
\newcommand{\tss}{$t_{ss}$}
\newcommand{\teff}{T$_{\rm eff}$}
\newcommand{\twch}{$^{12}$CH}
\begin{document}
\twocolumn[
\title{The s-process in metal-poor stars: Abundances for 22 neutron-capture
elements in CS 31062-050}
\author{Jennifer A. Johnson}
\affil{Dominion Astrophysical Observatory, Herzberg Institute of Astrophysics, National Research Council of Canada, 5071 West Saanich Rd., Victoria, BC V9E 2E7, Canada}
\and
\author{Michael Bolte}
\affil{UCO/Lick Observatory, University of California, Santa Cruz, CA 95064}
\begin{abstract}
The CH star CS 31062-050 ($\lbrack$Fe/H$\rbrack=-2.42$) is one of the most
useful stars yet discovered for evaluating the s-process in metal-poor
stars. It is very abundant in heavy elements
(e.g., $\lbrack$La/Fe$\rbrack=2.2$), and its relatively cool temperature
and low gravity mean that there are many lines of interesting elements
present in the spectrum. We measured the abundances of 22 elements with
Z$\geq$29, including the rarely measured Lu and Pd. We derive an
upper limit on the Th abundance as well. The abundances in
CS 31062-050 show a similar pattern to many other metal-poor CH
stars: high $\lbrack$Pb/Fe$\rbrack$ and $\lbrack$Pb/La$\rbrack$ ratios, low 
$\lbrack$Y/La$\rbrack$ ratios and high
$\lbrack$Eu/La$\rbrack$ values compared to the solar system s-process. However, the Th limit, 
with additional assumptions, is not consistent with the idea that the
excess Eu in CS 31062-050 is contributed by the r-process. In addition,
the observed $\lbrack$Eu/Tb$\rbrack$ cannot be explained by any ratio of solar-system
s-process and r-process abundances. We therefore argue that the abundance
pattern in CS 31062-050 is most likely the result of the s-process, and
we discuss possible modifications that could explain the non-solar-system
pattern observed.
\end{abstract}
\keywords{nuclear reactions, nucleosynthesis, abundances, stars:abundances --- stars: atmospheres --- stars: Population II}
]

\section{Introduction}
The sample of identified field stars with [Fe/H]\footnote{We use the usual notation 
[A/B]$\equiv {\rm log}_{10}(N_A/N_B)_* - {\rm log}_{10}(N_A/N_B)_{\odot}$ 
and log$\epsilon(\rm A) \equiv {\rm log}_{10}(N_A/N_H)+12.0$. 
A/B$\equiv N_A/N_B$.}$<-2.4$
has increased by more than an order of magnitude in the last decade.
  Because the elements in the atmospheres of these
stars have been produced in a small number of nucleosynthetic events,
abundance determinations can provide direct tests of model yields from
different nuclear processes.  Even in the early work on very
metal-poor stars it quickly became clear that there were subclasses of
objects with very high [heavy-element/Fe] ratios that could be traced
to specific nucleosynthetic origins (e.g., McWilliam \etal{} 1995).

The elements heavier than the iron peak are made through neutron capture
via two principal processes: the
r-process and the s-process (Burbidge \etal{} 1957). The r-process (for
rapid process) occurs when neutrons are added much more rapidly than
the $\beta$ decay times of the relevant nuclei.  The site or sites of
the r-process are not known, although suggestions include the
$\nu$-driven wind of Type II SNe (e.g., Woosley \& Hoffman 1992; Woosley \etal{} 1994) and the mergers of neutron stars (e.g., Lattimer \& Schramm 1974; Rosswog
\etal{} 2000). The
s-process occurs when neutrons are added more slowly and $\beta$
decays, changing neutrons to protons, keep the nuclei from straying
far from the valley of $\beta$ stability. The He intershell in 
asymptotic giant branch (AGB) 
stars is the site of the s-process as well as of C production.
C and heavy elements
 are brought to the surface of the AGB stars during the ``third dredgeup''.
Since the r-process produces very neutron-rich nuclei
initially, the r-process reaches
the neutron magic numbers with considerably fewer protons than the
s-process, and, as a result, these two processes produce abundance
peaks at different atomic weights. As a result, when the 
solar-system total
abundances (\tss) are separated into contributions from the s-process
(\sss) and the r-process (\rss) (e.g., K\"appeler, Beer, \& Wisshak, 1989;
Arlandini \etal{} 1999),
 some elements are mostly contributed by the r-process,
such as Eu, and some by the s-process, such as Ba and La. Therefore Eu
is commonly referred to as an ``r-process element'' and Ba and La as
``s-process elements''.  The [Eu/Ba] and the [Eu/La] values are used
to estimate the ratio of r-process to s-process contributions to the
heavy element abundances in a star, increasing as the r-process
fraction increases.  Despite the nomenclature, it is important to 
remember that all neutron-capture elements lighter than Z=84
are made in both processes (Clayton \& Rassbach 1967). 
Th (Z=90) and U (Z=92) can only be made in the r-process.

The surveys of metal-poor stars have uncovered stars 
rich in C and s-process elements and stars rich in r-process elements.
These have been the subject of many follow-up studies, because of the
insight they can provide on the production of the heavy elements in the
early Galaxy.

\subsection{The very metal-poor CH stars}

Metal-poor stars with enhanced abundances of C and s-process elements
are called CH stars.  In an
extensive survey, McClure (1984) and McClure \& Woodsworth (1990)
showed that all of the CH stars in their sample were members
of binary systems with orbital parameters consistent with a white dwarf
secondary. The explanation for the classic CH stars is that they
result from the transfer of C, N, and s-process material produced 
in an AGB companion which
is now a white dwarf.  The abundance patterns in CH stars for the heavy
elements are therefore a very accessible means of empirically deriving
s-process yields and for inferring the structure of, and physical 
conditions in, AGB
stars.
The traditional indicators of s-process material are super-solar
[Ba/Fe] (the Pop I version of CH stars are often referred to as barium stars)
and subsolar [Eu/Ba]. 

Theory predicts that s-process nucleosynthesis will depend
on the initial metallicity of the AGB star and on its mass (Gallino
\etal{} 1998; Goriely \& Mowlavi 2000; Busso \etal{} 2001).  Busso \etal{}  reported s-process
yields as a function of initial [Fe/H] and for AGB stars with 1.5 and
3.0M$_\odot$ and made comparisons with the available observational
data. The basic metallicity dependence is a tilting of the s-process
products toward heavier elements with decreasing [Fe/H] of the host AGB
star.  The prediction that $^{208}$Pb will have a particularly strong
excess has been verified for CS 22183-015 (Johnson \& Bolte 2002a), HE
0024-2523 (Lucatello \etal{} 2003) and many of the stars studied by Van
Eck \etal{} (2001, 2003) and Aoki \etal{} (2002) (A02). On the other
hand, Aoki \etal (2000, 2001) present the analysis of two metal-poor,
s-process-rich stars, LP 625-44 and LP 706-7, which have [Pb/Ba] $\sim$
0.

\subsection{Very metal-poor r-process-rich stars}
There is a second class of neutron-capture-rich very metal-poor
stars in which the abundance pattern of the heavy elements more closely
follows that inferred for the r-process elements in the solar system.
CS 22892-052 (Sneden \etal{} 1996, 2000), HD 115444 (Westin \etal{} 2000)
and CS 31082-001 (Cayrel \etal{} 2001) are the best studied members
of this class. Studies have showed a remarkable similarity
in the abundance ratios for elements between Ba (Z$=56$) and Hf (Z$=72$),
(c.f. Truran \etal{} 2002)
although in the 
lighter and heavier
r-process element peaks there is considerable star-to-star scatter
in abundance ratios (e.g., Sneden \etal{} 2000; Johnson \& Bolte 2002b;
Hill \etal{} 2002). Some r-process-rich stars, in particular CS22892-052,
are also C-rich. The source of that C is unknown. Finally, we note that
very metal-poor stars with low [neutron-capture/Fe] ratios have abundance
 patterns between Ba and Hf that agree with \rss{} (Sneden
\& Parthsarathy 1983; Gilroy \etal{} 1988; Johnson \&
Bolte 2001).

\subsection{Some puzzles and challenges} 

As more detailed studies of very metal-poor CH stars became available,
some stars and elements did not fit neatly into the picture described
above. Despite having lower [Eu/Ba] than the
r-process-element-rich  stars, in some CH stars [Eu/Ba] is higher
than that \sss{} or than predicted for the metal-poor s-process.
Hill \etal{} (2000)
 analyzed spectra of two CH stars, CS 22948-027 and CS 29497-034, 
and concluded that the observed abundances could not be fit 
by either a scaled solar-system s-process (\sss) or scaled
solar-system r-process (\rss), but instead reflected enrichment by
both processes. 

A02 and Johnson \& Bolte
(2002a) also noted a large spread in [Eu/Ba] for other 
CH stars but suggested that the abundance pattern seen in
these CH stars was due solely to the s-process, albeit one that
produces a varying Eu/Ba ratio.  A large percentage of CH stars have
Eu/Ba ratios larger than \sss. With the Hill \etal{} interpretation,
this would suggest that a number of s-process-rich stars are also
r-process rich. Some fraction of non-CH field stars are Eu-rich, so it would
not be surprising to find some stars that began as r-process-rich and also
were polluted by an AGB companion later in their lives. However, while
seven of 32 non-CH stars in McWilliam \etal{} (1995) have [Eu/Fe]\gtsim 0.5,
six of eight CH stars in A02 have such high Eu. 
So it appears that r-process enrichment cannot be the solution for {\it all}
the high Eu CH stars (but see below).

Recently, Cohen \etal{} found an extreme example of non-\sss{} ratios
in the CH star HE 2148-1247. The measured Ba/Eu ratio was $\sim 100$
while theoretical calculations from Arlandini \etal{} (1999), for
example, give Ba/Eu $\sim$ 640. They favored the addition of
r-processed material as well as s-processed material to HE 2148-1247.
Qian \& Wasserburg \etal{} (2003), in a companion paper, proposed an
intriguing theory for the creation of such ``s+r''-process stars. First
some s-processed material is accreted from the AGB companion, which
turns into a white dwarf.
 Later in the evolution of the system, the white dwarf accretes
matter from the polluted star and suffers an
accretion-induced collapse (AIC) to a neutron star. 
The $\nu$-driven wind produces an
r-process, which also pollutes the companion, but since the white dwarf
lacks an H or He envelope, lighter elements, such as Fe, are not
manufactured. Therefore, the remaining star is r-process {\it and}
s-process-rich. The AIC could potentially deliver a strong kick to the
neutron star, 
which could explain why some CH stars have recently been
shown not to be members of binaries (Preston and Sneden 2001; Hill
\etal{} 2000).  Because the production of the s-process in the AGB star
and the production of the r-process in the AIC are connected to the
same binary companion, this alleviates some of the concern expressed by
A02 and Johnson \& Bolte (2002) about the high frequency of potential
s$+$r stars. However, at least 50\% of the 
very metal-poor CH stars would
need to be s$+$r stars if that is the explanation for the elevated Eu/Ba
ratios (see Figure 13 in Cohen \etal{} (2003)). As Qian \& Wasserburg
point out, the frequency of AIC events and the parameters necessary to
create them are unknown at this time, but the large fraction of very
metal-poor CH stars with high Eu implies that the ability to pollute
a companion as a AGB star must be tightly correlated with an r-producing
AIC event. Another possible problem these authors mention is the still
uncertain nucleosynthesis in AIC, which may or may not produce the r-process.

If we hope to understand the origin of neutron-capture elements in CH stars,
we need to measure the abundances of as many heavy elements as possible
to see if they are consistent with an s$+$r, or s-only or r-only scenario.
In this paper,  we present results for another CH star. Because of
the combination
of large abundance enhancements and 
atmospheric parameters in this star, we can measure 
accurate abundances for a number of elements that have
been infrequently studied in CH stars. These include Pd, Tb, Ho, Tm
and Lu. We have also been able to put interesting
limits on the abundance of Th.

\section{Observations}
CS 31062-050 was selected from a survey of metal-poor star candidates that
we have undertaken with the Keck 2 telescope and the intermediate-resolution
echelle spectrometer ESI (Sheinis \etal{} 2002).
The goal of the ESI program is to obtain
the abundances of Fe, Ti, Mg, Ca, Ba and in some cases Eu for very metal-poor 
star candidates and to identify interesting targets
for follow-up at higher resolution and bluer wavelengths with Keck 1 and 
HIRES (Vogt \etal{} 1994).  
CS 31062-050 was identified as a metal-poor star candidate
 by the UBV photometry of
Norris, Ryan \& Beers (1999).
An ESI spectrum was obtained on 29 August 2000. A single 900s exposure gave  
a high S/N ($> 100$) spectrum at R$\sim 7000$ from 3900\AA{} to 
1$\mu$m.  The high carbon abundance of CS 31062-050 was immediately evident from
the strength of the G-band. The ESI spectrum also revealed
strong \ion{Ba}{2} and \ion{Eu}{2} lines. We took higher-resolution spectra
with HIRES on Keck I on 25 and 26 August 2001. A 0\Sec86 slit
gave R$\sim$45,000. A blue setting which covered
the wavelength 3190\AA--4710\AA{} was used for three observations of
1800s each. The second grating setting covered 3730\AA--5275\AA{} and was also used
for three observations of 1800s. Exposures of a quartz lamp were used for 
flatfielding and of a 
ThAr lamp for wavelength calibration. The frame processing was done in IRAF\footnote{IRAF is distributed by the National Optical Astronomy Observatories,
which are operated by the Association of Universities for Research
in Astronomy, Inc., under cooperative agreement with the National
Science Foundation}.
In addition to combining the three observations at each setting, we also
combined all six observations together where they covered 
the same wavelengths. Our S/N ranged from $\sim$ 10 at 3200 \AA{} to $\sim$80
at 4000\AA{} to $\sim$ 100 at 5250\AA. This star was also studied by A02, 
but at lower S/N and without the very blue wavelength coverage. They found
that this star was extremely C- and N-rich, with [C/Fe]$=2.0$ and [N/Fe]$=1.2$
We were able to measure the abundances of several neutron-capture 
elements not included in the A02 study, and these provide interesting
tests for theories of the origin of the heavy elements in the early Universe.

\section{Abundance Analysis}
MOOG (Sneden 1973) was used for the abundance analysis. We used
Kurucz\footnote{http://cfaku5.harvard.edu/} 
(2003) model atmospheres. Our \teff{} comes from the excitation
equilibrium of \ion{Fe}{1} lines. Deriving \teff{} from colors in carbon-rich
stars can be problematic because the presence of a large number of molecular
lines affects the observed colors. However Hill \etal{} (2000) found that their excitation equilibrium
temperature was consistent with temperatures derived from colors when
they included the effects of the molecular lines on the atmospheres
and on the colors. 
Our gravity comes from ionization equilibrium
of \ion{Fe}{1}/\ion{Fe}{2}. The validity of this assumption in view of possible
NLTE effects on \ion{Fe}{1} lines has been the subject of much recent
discussion (see e.g., Lucatello \etal{} 2003; Johnson 2002), but
the abundances of most of the heavy elements are affected in the same
manner by a change in log $g$. 
The microturbulent velocity ($\xi$) was determined
by requiring \ion{Fe}{1} lines of different equivalent widths (EWs) 
to give the same [Fe/H]. 
Our final atmospheric parameters
are \teff=5500K, log $g$=2.70, $\xi$=1.3 km/s and [Fe/H]$_{mod}=-$2.30. 
These agree well with the parameters adopted by A02
for this same star (\teff=5600K, log $g$=3.00 and $\xi$=1.3 km/s).
This implies too low a luminosity for this star to be an AGB star. We also
checked for the presence of Tc lines at 4049\AA{} and 4088\AA{} using the atomic
parameters from Vanture \etal{} (1991). No Tc was detected, which supports our
conclusion that this not a self-polluting AGB star, 
although large Tc enhancements would be required for any
lines to be visible. Finally, Aoki \etal{} (2003) have found radial
velocity variations in this star. We conclude that CS 31062-050 is a CH star.

We selected a number of lines with good \gfvalues. Whenever possible,
we used recent laboratory measurements.  Our choices for \gfvalues{}
for the heavy elements are summarized in the Appendix. Hyperfine
splitting (HFS) and isotopic splitting (IS) are important for many
lines of the heavy elements, and we included them for all the elements
with HFS and IS constants for the relevant levels in the
literature. These sources are discussed in the Appendix as well. For
elements lighter than Cu, we used the linelists of Johnson (2002).

The large enhancements of neutron-capture elements meant that we could
measure many lines of heavy elements, including some not
identified in the solar atlas of Moore, Minnaert \& Houtgaust (1966)
or in previous abundance
analyses of metal-poor stars. To find other lines that would be useful
for abundance analysis, we created lists of all the lines from our sources
of \gfvalues. We then predicted their EWs using MOOG. For lines with 
EWs $> 5$m\AA{} (slightly more for the lower S/N regions), 
we synthesized the region. Our starting point was the Kurucz \& Bell (1995) (KUR95)
line lists, which contained many rare earth lines that
are the true contaminants in this situation, but were unimportant in the Sun
and not included in Moore \etal.
For lines we wanted to use for abundance analysis, we checked for blends
from other lines and only used a line if it were essentially unblended
or if the other line contributing to the blend had a recent, laboratory-based
 \gfvalue. 
We also synthesized the regions in three metal-poor stars: HD 122563 
(neutron-capture and C-poor) (e.g., Westin \etal{} 2000), CS 22183-015 (neutron-capture and C-rich) (Johnson \& Bolte 2002), and
CS 22957-027 (neutron-capture-poor and C-rich) (e.g., Norris, Ryan, \& Beers 1997) 
to check that our linelists adequately accounted 
for contamination from atomic and molecular lines.
Our EWs and atomic parameters are in Table 1. We compared our EWs to those of A02 for the 22 lines we had in common. We found an average offset of 
$-0.8$m\AA{}
with an r.m.s. scatter of 5.9 m\AA, indicating excellent agreement
between the two studies.
Below we include some notes on three of the most 
interesting or uncertain abundances.

\subsection{Palladium}
A rarely measured element is the intermediate-mass neutron-capture
element Pd. With A$\sim$130, it lies between the better-studied
regions near Sr (A$\sim$90) and Ba (A$\sim$160). When this region
was first explored in metal-poor r-process-rich stars, unexpected
deviations from \rss were found (Sneden \etal{} 2000
; Johnson \& Bolte 2002b). 
Of all the elements in this mass region, 
we could measure one line of Pd. Our synthesis is shown in Figure 1.
\begin{figure}[htb]
\begin{center}
\includegraphics[width=3.0in,angle=0]{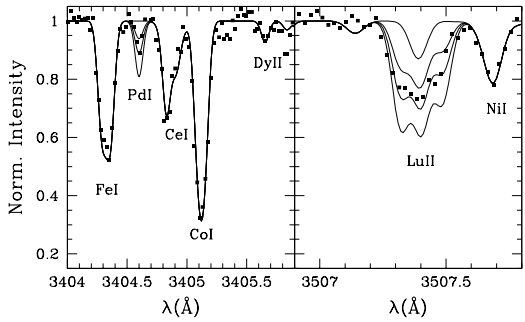}
\caption{Spectral synthesis of (left) \ion{Pd}{1} and (right) \ion{Lu}{2} in
CS 31062-050. The solid squares are the data. The solid lines
represent, in order of increasing strength and relative to our
adopted log $\epsilon = -\infty$, $-0.3$ dex, 0.0 dex, and $+0.3$ dex. 
Several lines in these regions of the spectrum have uncertain \gfvalues{}
and we have increased the \gfvalue{} for \ion{Fe}{1} on the left and \ion{Ni}{1} on
the right above that in KUR95 to fit the spectrum.}
\end{center}
\end{figure}
The \gfvalue{} is from Bi\'emont \etal{} (1982) and the line list
is from Johnson \& Bolte (2002b). Unfortunately, none of the odd
elements in this mass region, such as Ag, could be measured because their
lines are weaker than that of Pd and they are in regions of the
lower S/N. 

\subsection{Lutetium} We were able to detect the 3507.4\AA{} line of \ion{Lu}{2} in
\csstar. It has a very accurate \gfvalue{} and HFS constants from the 
study by Den Hartog \etal{} (1998). 
It is blended in CS 31062-050 with an \ion{Fe}{2} line.
This \ion{Fe}{2} line has a \gfvalue{} from Moity (1983).  
Figure 1 shows our synthesis of the \ion{Lu}{2} line. 
The hyperfine splitting of the Lu line is significant enough to
noticeably broaden the line. Figure 1 shows that no
amount of absorption from \ion{Fe}{2} will account for the profile of the
observed line, while the three components of the \ion{Lu}{2} line 
are a good match when the S/N is taken into account.

\subsection{Thorium} 

A possible
detection of the r-process-only element Th 
was one of the reasons that Cohen \etal{} (2003) favored an
s+r scenario. 
Unfortunately, in CS 31062-050 the strongest Th line at $\lambda$4019\AA{}
is blended with a \thch{} line (Norris, Ryan,
\& Beers 1997). 
Because we 
could only set large upper limits ([Th/Fe]$<3.0$) using the non-detection of
other, non-blended Th lines in our spectra,  we 
decided to use the line at 4019\AA. The linelist from
Johnson \& Bolte (2001) was used, except our \gfvalue{} for the \ion{Th}{2} line 
was revised up by 0.05 dex to reflect the recent laboratory measurement of
the oscillator strength by Nilsson \etal{} (2002). The \ion{Ce}{2} line noted
by Sneden \etal{} (1996), which
in our previous work was an unimportant contributor to the absorption,
is much more important in this s-process-rich star. 
Our fit to the left of the Th line was improved if we shifted the
\ion{Ce}{2} line by 0.02\AA{} to the red from its
position in Johnson \& Bolte (2001) and its \gfvalue{} decreased by 0.1 dex. 
This was a cosmetic change and does not affect
the Th abundance determination. In an attempt to test
the goodness of the fit of our synthesis spectra with different Th abundances,
we did a $\chi^2$ test of the wavelength region of 4019.08--4019.25\AA. 
The errors in our
data points were assumed to be solely due to the Poisson noise in the counts,
The smoothing of the synthesis was set by other lines in the region
and not allowed to vary. The C abundance was set to match the \twch{} lines
at 4020\AA{} ([C/Fe]=1.82), and a \twch/\thch{} value of 12 was used.
We find a 3$\sigma$ limit of log $\epsilon (Th)=-0.85$ and a 5$\sigma$ 
limit of log $\epsilon (Th) = -0.70$. Our syntheses for these limits
are shown in Figure 2.
\begin{figure}[htb]
\begin{center}
\includegraphics[width=3.0in,angle=0]{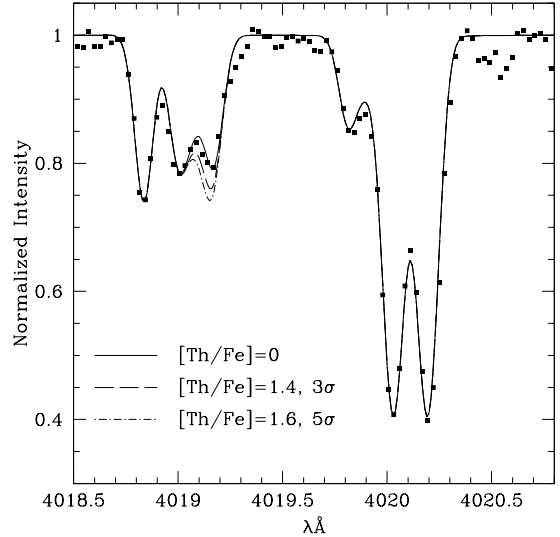}
\caption{Spectral synthesis of the Th line at 4019\AA. The solid squares
are the data. At [Th/Fe]=0, the Th line is
so weak that our synthesis is indistinguishable from a synthesis with
no Th.}
\end{center}
\end{figure}

\section{Results}
Our abundances and errors
are in Table 2. The solar values are adopted from the meteoritic abundances
in Grevesse, Noels, \& Sauval (1996). 
Table 2 also includes a comparison with the A02 abundances.
Our lower \teff{} and gravity mean that we have lower log$\epsilon$ in
general for the singly-ionized rare earth species, but the offset from
A02 is fairly constant. The abundances of the elements between Mg and Ni are in good agreement
with metal-poor field stars that are not C-rich or heavy-element-rich
studied with the
same linelist by Johnson (2002). The ``$\alpha$''-elements (Mg, Ca, and Ti) 
have supersolar [X/Fe] ratios. The iron-peak elements track Fe
closely with the exception of Mn, whose depletion relative to iron in
metal-poor stars was first noted by McWilliam \etal{} (1995).
Therefore, any explanation of the over-enhancements of 
the neutron-capture elements cannot change the lighter elements away from
the standard halo patterns. [Cu/Fe] and [Zn/Fe] in CS 31062-050 agree with
the trends seen in the majority of metal-poor stars (Sneden, Gratton, \& Crocker 1991; Primas \etal{} 2000).

\begin{deluxetable}{lrrrrcr}
\tablenum{2}
\tablewidth{0pt}
\tablecaption{Abundances}
\tablehead{
\colhead{Ion} & \colhead{log $\epsilon$} & 
\colhead{[X/Fe]\tablenotemark{a}} & \colhead{$\sigma$} & \colhead{$\sigma_{tot}$} &
\colhead{N$_{lines}$} & \colhead{$\Delta$ A02\tablenotemark{b}} 
}
\startdata
\ion{Mg}{1} & 6.01 & 0.84 & 0.20 & 0.21 & 1 & \\
\ion{Ca}{1} & 4.43 & 0.49  & 0.10  &0.09 & 2 & \\
\ion{Ti}{1} & 2.85 & 0.32 & 0.25 & 0.13 & 9 &\\
\ion{Ti}{2} & 2.86 & 0.33 & 0.10 & 0.23 &  4 &\\
\ion{V}{2} & 1.44 & $-$0.17& 0.14  &0.18 & 2 &\\
\ion{Cr}{1} & 3.36 &0.08 & 0.19 & 0.15 & 3 &\\
\ion{Mn}{1} & 2.69 & $-$0.43& 0.12 &0.12 & 4 &\\
\ion{Fe}{1} & 5.09 & $-$2.41 & 0.11 &0.13 & 46 & $-$0.10 \\
\ion{Fe}{2} & 5.10 & $-$2.40 &0.13 &0.15 &10 & $-$0.07\\
\ion{Co}{1} & 2.60 & 0.10  & 0.20 & 0.23& 1 &\\
\ion{Ni}{1} & 3.68 & $-$0.16 & 0.17 & 0.21 & 11 &\\ 
\ion{Cu}{1} & 0.69 & $-$1.19& 0.20 & 0.18 & 2 &\\
\ion{Zn}{1} & 2.32 & 0.06 & 0.20 & 0.23 & 1 & \\
\ion{Y}{2} & 0.30 & 0.48 &0.10 & 0.22  & 14 &\\
\ion{Zr}{2} & 1.05 & 0.85 & 0.14 & 0.19 &9 & $-$0.26\\
\ion{Pd}{1} & 0.27 & 0.98 & 0.20 & 0.24 & 1 & \\ 
\ion{Ba}{2} & 2.61 & 2.80& 0.20 & 0.19 & 3 & 0.40\\
\ion{La}{2} & 0.93 & 2.12 & 0.12 & 0.19 & 25 & $-$0.41\\
\ion{Ce}{2} & 1.24 & 2.02 & 0.16 &0.18 & 36 & $-$0.17\\
\ion{Pr}{2} & 0.13 & 1.74 & 0.11 & 0.16 & 10&  \\
\ion{Nd}{2} & 1.07 & 1.99 & 0.18 & 0.22 & 39 & $-$0.34 \\ 
\ion{Sm}{2} & 0.53 & 1.96 & 0.14 & 0.18 & 12 & $-$0.28\\
\ion{Eu}{2} & $-$0.07 & 1.79 & 0.07 & 0.17 & 5 & $-$0.14 \\
\ion{Gd}{2} &0.51 & 1.83 & 0.12 & 0.19 & 24 & \\
\ion{Tb}{2} & $-$0.56 & 1.50 & 0.10 & 0.16 & 4 &\\
\ion{Dy}{2} & 0.39 & 1.63 & 0.07 & 0.23 & 20 & $-$0.54 \\
\ion{Ho}{2} & $-$0.40 & 1.55 & 0.16 & 0.18 & 3 & \\
\ion{Er}{2} & 0.78 & 2.22 & 0.13 & 0.26 & 15 & \\
\ion{Tm}{2} &$-$0.29 & 1.97 & 0.10 & 0.19 & 4 & \\
\ion{Yb}{2} & 0.76 & 2.21 & 0.20 & 0.30 & 1 &\\
\ion{Lu}{2} & $-$0.10 & 2.18 & 0.20 & 0.26 & 1 &\\
\ion{Hf}{2} & 0.67 & 2.33 & 0.10 & 0.20 & 11 &\\
\ion{Pb}{1} & 2.46 & 2.81  & 0.15 & 0.15 & 2 & $-$0.14 \\  
\enddata
\tablenotetext{a}{[X/Fe] for all elements except Fe, where [Fe/H] is given}
\tablenotetext{b}{log$\epsilon_{this{}study}-$log$\epsilon_{A02}$}
\end{deluxetable}

In Figure 3, we plot the abundances of CS 31062-050 as well as 
\rss, \sss, and \tss{} scaled to match the abundance of
La. We use La rather than the more traditional Ba because of the large 
uncertainties in measuring the Ba abundance (see Appendix).
\begin{figure}[htb]
\begin{center}
\includegraphics[width=2.5in,angle=270]{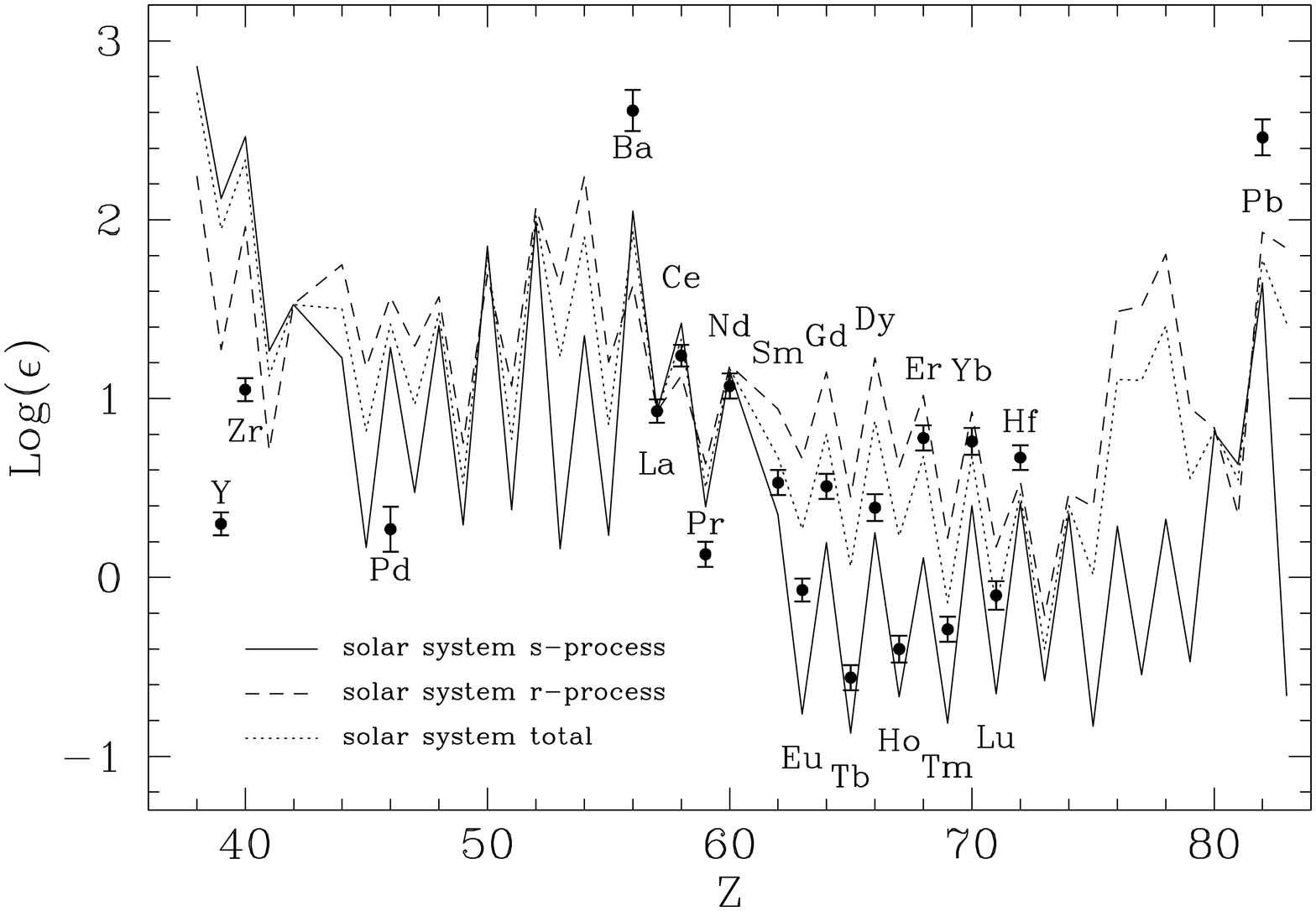}
\caption{The abundance in CS 31062-050 compared with \sss, \rss, and
\tss. In all cases the solar abundances were scaled to match
the La abundance of CS 31062-050. No solar abundance pattern is a good match
to the data, even in a limited Z range.}
\end{center}
\end{figure}
It is clear that none of \tss{}, \sss{} or \rss{} is a good match to the data.
In several respects, \sss{} is the closest, particularly for explaining
elements like La, Ce, Pr, Nd, Tb, Dy, and Ho. However,
while Y and Zr are enhanced relative to the majority of metal-poor stars,
they are not as abundant as a simple scaling of \sss{} implies. Pd is
also $\sim$ 1 dex below the solar system lines.  Pb, on the other hand
is above the solar system lines with [Pb/Fe]=2.81 and [Pb/La]=0.69.
Although this value is not as extreme as CS 22183-015 (Johnson \&
Bolte 2002a) or the stars in Van Eck \etal (2001), it is supersolar.  
 
We can understand this tilt toward heavier nuclei as the result of a
high neutron-to-seed ratio (Gallino \etal{} 1998). Figure 4 shows the
abundances in CS 31062-050 compared with the results of an s-process
in a 1.5$M_{\odot}$ AGB star with [Fe/H]$=-2.3$.
(R. Gallino, private communication). 
\begin{figure}[htb]
\begin{center}
\includegraphics[width=3.0in,angle=0]{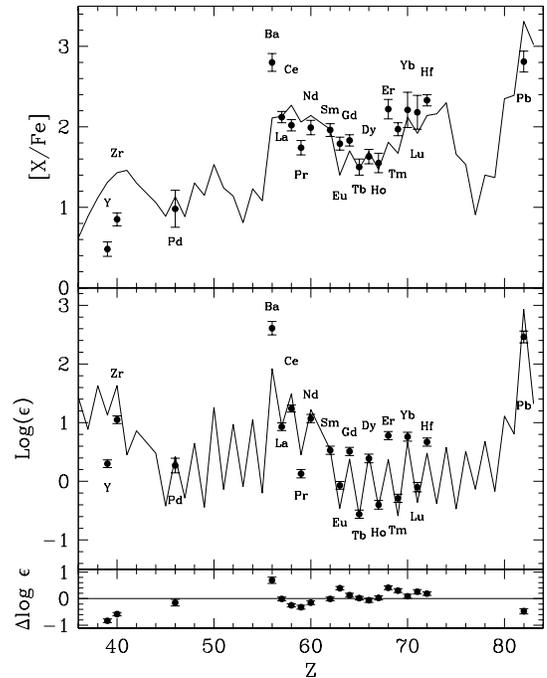}
\caption{The abundance in CS 31062-050 (filled circles) 
compared with an s-process calculation for a metal-poor AGB star. 
The top panel shows [X/Fe], which highlights the rise in overabundances
of the neutron-capture elements as Z increases. The middle panel shows
log $\epsilon$ vs. Z, which shows that
the metal-poor s-process model does a better
job of fitting the data than scaled solar system abundance patterns, although
Eu and Er are both underpredicted. The difference between the model and the
data is shown in the bottom panel.}
\end{center}
\end{figure}
The top panel of Figure 4 shows the steep rise in [El/Fe] as Z increases,
which is matched much better by the metal-poor s-process than by
\sss, although  Y and Pb are not simultaneously fit.
However, the agreement with Pd, Sm, Tb, Dy and Ho is very encouraging and
a marked improvement over Figure 3.
The bias to heavier nuclei
can also be quantified using the
light s-process to heavy s-process ratio [ls/hs]. The value for CS
31062-050 (using the elements suggested by Busso \etal{}, 2001) is
1.39, in good agreement with predictions for the s-process in
[Fe/H]$\sim -2.5$ AGB stars and with other observations of CH stars
(Busso \etal{} 2001). This model predicts no appreciable production of 
Cu or Zn.
The low Cu and Zn observed in CS 31062-050 (Table 2)
are also consistent with the conclusions of Matteucci \etal{} (1993)
that Cu is primarily made in either explosive nucleosynthesis or the
weak s-process and Zn in explosive nucleosynthesis, 
because they are clearly not made in the AGB star that polluted CS 31062-050.

The most glaring misfits in the Z$\geq 56$ range are the abundances of Eu
and Er (and, with larger error bars, Ba). 
It was a similar observation, in
particular the high Eu abundance, that prompted the suggestions of
Hill \etal{} and Cohen \etal{} (2003) that CH stars with this characteristic
were rich in
both the s- and r-process. We now examine this question in more
detail.

\section{Sources for the heavy elements in CS 31062-050}

There are now a number of very metal-poor CH stars with high-resolution
abundance analyses. A surprising number of them have [Eu/La] too high to
be consistent with \sss{} (Figure 5). Since Eu is predominately
produced in the r-process in the solar system, the r-process is an
clear choice for the source of Eu in CS 31062-050.
If the r-process contributes an appreciable amount to the heavy elements 
then we would expect relatively high abundances of 
elements mostly or only produced in the r-process. Unfortunately, we
could only obtain uninteresting upper limits for elements in the 
third r-process peak such as Os and Ir. However, the limit for Th,
an r-only element, is more interesting.
\begin{figure}[htb]
\begin{center}
\includegraphics[width=3.0in,angle=0]{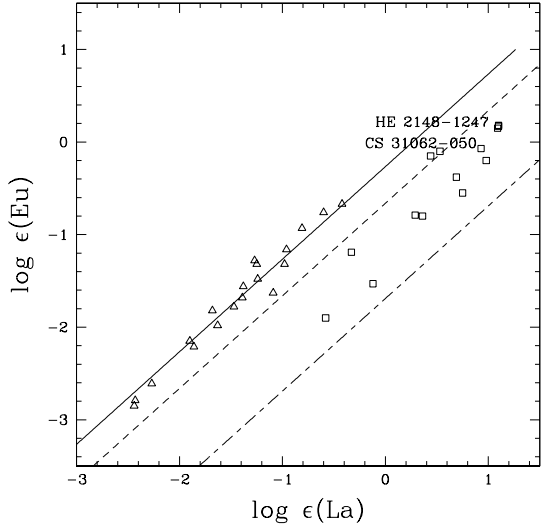}
\caption{Log $\epsilon$(Eu) vs. log $\epsilon$(La) for a sample of
CH (squares) and non-CH stars (triangles) with [Fe/H]$<-2.0$. The lines represent
an \rss{} ratio (solid), \sss{} ratio (dot-dash) and \tss{} ratio (dashed)
of La to Eu. The non-CH stars are identified by their adherence to the 
\rss{} line and have a wide range of Eu (and C) abundances. The CH stars 
have a wide range of La/Eu ratios and large enhancements in both elements. }
\end{center}
\end{figure}
\subsection{Thorium abundance?}
We found an 3$\sigma$ upper limit to the abundance of
Th of log$\epsilon=-0.85$. What does this tell us about the possible 
contributions of the r-process to the lighter neutron-capture elements,
such as Eu?
We cannot definitely predict the
expected Eu abundance in CS 31062-050 from the Th abundance
since the ratio between Eu and the initial Th value produced in the
r-process is not always constant. 
One theoretical calculation 
predicts Th/Eu=0.496 (Cowan \etal{} 1999),
and many stars show that to be a reasonable value (Sneden \etal{} 
2000; Johnson \& Bolte 2001; Cowan \etal{} 2002).
While CS 31082-001 (Hill \etal{} 2002)
clearly disagrees with this Th/Eu ratio, its implied Th/Eu$_{initial}$ is
higher than the value we use here, which would only decrease the possible 
contribution
to the r-process to the Eu abundance derived from our Th limit. If we use Th/Eu=0.496 and an age for CS
31062-050 of 14 Gyr, we find that at most 66\% of our Eu abundance could
come from the r-process. We can now perform the same calculation for
the s-process contribution, using the observed La/Eu ratio and the 
solar system s-process ratio from Arlandini \etal{} (1999). 
Here we find that the s-process only contributes 23\% of the Eu, leaving
at least 10\% of the Eu without a source. Either our assumed s-process or
r-process ratios are incorrect.
Clearly, more r-process-rich stars with Th determinations will eventually
better constraint the possible (Th/Eu)$_{initial}$ range, but based on
current data, our upper-limit on Th argues against the r-process as the
solution to the abundance pattern in CS 31062-050. 

\subsection{[Eu/Tb]}
The ratio of the predominately r-process elements Eu and Tb provide
even stronger evidence that no combination of \rss{} and \sss{} or current
models for the metal-poor s-process can fully explain the
abundnaces measured in CS 31062-050. 
According to Arlandini \etal{} (1999), 94.2\% of the solar
system Eu is due to the r-process as is 92.8\% of the solar system Tb.
Therefore, only a small variation in the ratio of Eu/Tb can be produced if we 
confine ourselves
to the solar system patterns. A pure r-process results in Eu/Tb
of 1.52, and a pure s-process results in Eu/Tb of 0.625.
 Neither pure \sss{} nor pure \rss{} nor any combination of those two can create the
Eu/Tb=3.1 that we measure in CS 31062-050. These predictions for \rss{} for
Eu and Tb are the result of subtracting \sss{} from the total solar
abundances, so the predictions for the two processes are correlated. 
In this case, given the overwhelming proportions of Eu and Tb produced
in the r-process in the solar system, changing \sss{} will not affect
\rss{} significantly, and so our argument that the r-process is not
responsible is robust against changes in 
\sss. In addition, the Busso \etal{} (2001) 
metal-poor s-process models predict similar abundances of Eu
and Tb. Therefore any combination of these models and \rss{} also will 
fail to reproduce the high Eu/Tb value. 
\begin{figure}[htb]
\begin{center}
\includegraphics[width=1.8in,angle=270]{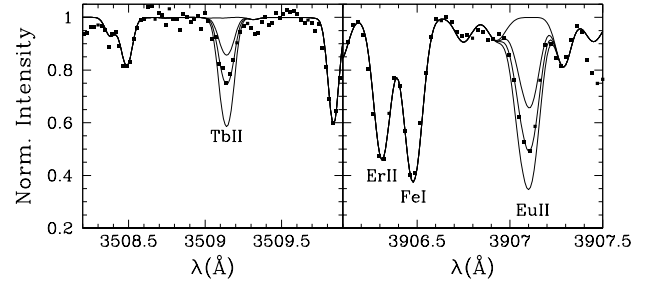}
\caption{Spectral synthesis of a line of (left) \ion{Tb}{2} and (right) 
\ion{Eu}{2}.
The solid squares are the data. The solid lines represent, in
order of increasing strength and relative to our adopted log $\epsilon =
-\infty$, $-0.3$ dex, 0.0 dex, and $+0.3$ dex.}
\end{center}
\end{figure}
Could observational error explain the large [Eu/Tb] ratio? 
We show the syntheses of one line of Tb and Eu in Figure 6. Our S/N
is good enough that the quality of our spectrum is not the problem.
The \gfvalues{} and HFS and IS constants for these two species are
based on laboratory studies. The line parameters are either the same as or
very similar to the values used by Sneden \etal{} (1996, 2002), Hill \etal{}
(2002) and Cowan \etal{} (2002) in their studies of extremely r-process-rich
stars. Therefore, while it is possible that our knowledge of the 
atomic parameters for these lines is in error, such a revision would
destroy the excellent agreement between the abundances in metal-poor
r-process-rich stars and \rss. We used the partition functions for
Eu and Tb from the 2002 release of MOOG\footnote{ftp verdi.as.utexas.edu}. We have not included any corrections due
to NLTE effects on the strong lines of EuII, but the work of Gehren 
\etal{} (2001) shows that this should be on the order of only 0.10 dex. 
HFS affects $^{151}$Eu more than $^{153}$Eu,
so if the isotopic ratio was different than the 50-50 split seen in the
solar system and predicted for the s-process (Arlandini \etal{} 1999),
the Eu abundance could be too high. However, even assuming that all the
Eu is in $^{151}$Eu reduces the Eu abundance by only 0.05 dex. We have
focused on Eu because we have HFS and IS information for its lines. Er is
also high relative to s-process models, but we could not find information
on the IS for our lines. Our largest Er EW is 64.4 m\AA, and
we see no trend of increasing abundance with increasing EW. So we 
believe our Er abundance is accurate, but would like the HFS and IS data to 
confirm this. Eu/Ho and Eu/Dy in CS 31062-050 are also higher than predicted
by the s-process models. Like Eu and Tb, Ho (93\%) and Dy (85\%) are mostly
produced by the r-process in the solar system and therefore a match
cannot be made to the observed ratios by adjusting contributions from \sss{}
and \rss.

Finally, we note that this study is
not the only one to find ratios that cannot be explained by
an s+r scenario. The Eu/Dy ratios for CS 22948-027 and CS 29497-034 
in Hill \etal{} (2000) are so high (0.74 and 1.32, respectively) that
they cannot be explained by a pure r-process (Eu/Dy=0.174), much less
a pure s-process (Eu/Dy=0.097) or any combination therefore. Unfortunately,
the Dy abundances in these two stars are based on only one line, so this
conclusion is not as robust as the Eu/Tb ratio in CS 31062-050. 
 Qian \& Wasserburg (2003) point out that the abundance of Gd in
HE 2148-1247 was too high to be fit by their best-fit single combination of 
\sss{} and \rss{}.

\subsection{The s-process and high Eu}
The results discussed in the previous sections are most economically
explained by an s-process that produces more Eu than currently
expected. Are there conditions in which an s-process produce such a 
high Eu/La ratio? To check this
possibility, we looked at the s-process calculations of Malaney (1987a).
These are parametric calculations, rather than calculations of the s-process
in a specific AGB star. Because we wish to see if slow neutron-capture
under any conditions will work, these are appropriate. Malaney (1987a) 
calculated abundances from Fe to Tl for a range of mean neutron exposures ($\tau_0$)
for two different neutron densities: N$_n=10^8$ and N$_n=10^{12}$. His
calculations are for a solar metallicity number of seeds, so we would
not expect his models to show the correct [ls/hs] metallicity dependence.
One of his models, $\tau_0$=0.05, N$_n=10^{12}$, was
a good match to the La/Eu ratio (Figure 7) and gives a reasonable
match to our measured abundance pattern from La to Hf.
\begin{figure}[htb]
\begin{center}
\includegraphics[width=3.0in,angle=0]{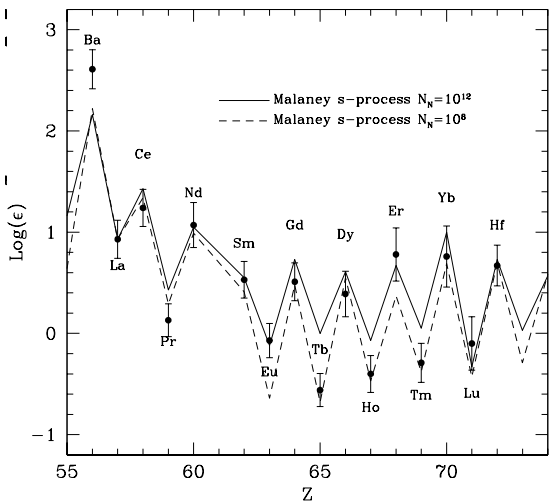}
\caption{Comparison of the abundances for CS 31062-050 (solid circles)
and a parametric s-process calculation of Malaney (1987a) with
$\tau_0$ and N$_n=10^{12}$ and N$_n=10^{8}$. The differences between
the two neutron densities is a result of the nuclear flow passing
through more neutron-rich nuclei if possible. These nuclei that have
smaller neutron-capture cross-sections, thus altering the relative
ratios of the elements (Malaney 1987b).}
\end{center}
\end{figure}  
Tb is a notable mismatch, and this neutron
density is much higher than expected for the s-process conditions in
an AGB star (Busso \etal{} 1999) in current models. In fact, Aoki
\etal{} (2003) derived constraints on the neutron density and the
temperature in CS 31062-050 by using isotopes of Eu that are sensitive 
to the branching point at $^{151}$Sm. They measure a $^{151}$Eu fraction
of 0.55, higher than the solar value of 0.48, which indicates a
neutron density of N$_n$=$10^7$-$10^9$. However, they make no predictions for
the absolute Eu abundance in CS 31062-050 in that paper. 
In addition to 
variations in physical conditions between AGB stars, 
the details of s-process branchings in the Eu region are 
still subject to theory and laboratory-based revisions. For example 
Best et al. (2001) report new ($n,\gamma$) cross sections for 
$^{152, 154}$Sm, $^{151, 153}$Eu and $^{164, 170}$Sm 
based on laboratory measurements irradiating Sm, Eu and Er samples with 
neutrons with a quasistellar energy spectrum. Differences with previous 
measurements are in some cases as large as 50\%, although there is a clear 
convergence with measurements by different groups since the 
1990s show agreement at the 20\% level. These authors also point out the 
uncertainties in the ``stellar enhancement'' corrections required to 
account for cross section and $\beta$-decay rate differences between 
laboratory and stellar interior conditions (due to the populations of 
low-lying excited states in many of the important nuclei). In the end,
a successful model will need to fit both the Eu abundance and the
Eu isotope fractions to explain the origin of Eu in this star using
appropriate cross sections and physical conditions. 

\section{Summary}
We have determined the abundances of 20 neutron-capture elements 
in CS 31062-050. This is a C-rich star with several ratios of neutron-rich
elements that match models of the metal-poor s-process.
However, several well-measured heavy element ratios, 
most notably Eu/La, are inconsistent with any s-process predictions
and the two ratios Eu/Tb and Eu/Dy cannot be explained by any
combination of any s-process predictions and r-process predictions.
High Eu/La (in the context of the s-process) has been seen before in
CH stars leading some authors to suggest CH stars that are
both s- and r-process-element rich. For CS 31062-050, our upper limit
to the Th abundance argues against the r-process as the complete explanation, 
at least in this star. 
The Eu/Tb and Eu/Dy ratios
suggests that our understanding of the s- or r-process yields
are incomplete for at least some of these elements. An s-process that
produces variable amounts of Eu in excess of the current models
could eliminate the most glaring differences between observations
and models. 

\acknowledgements
Data presented herein were obtained at
the W.M. Keck Observatory, which is operated as a scientific partnership
among the California Institute of Technology, the University of California,
and the National Aeronautics and Space Administration. The Observatory
was made possible by the generous financial support of the W.M. Keck
Foundation. Many thanks to Roberto Gallino for sharing the results of the s-process
predictions in advance of publication and for many helpful insights
on the workings of the s-process. We would also 
like to thank Sara Lucatello, Falk Herwig, and Jim Hesser for their
comments on drafts of this paper. We gratefully acknowledge support
from National Science Foundation grant AST-0098617.

\appendix{Atomic Parameters, Oscillator Strengths and Linelists}

{\it Copper:} The usual Cu lines at $\lambda 5105$\AA{} and
$\lambda 5782$\AA{} were undetectable, but we were able to measure the
resonance lines at $\lambda$3273\AA{} and $\lambda$3247\AA{}.  Because
of the HFS and IS of these lines, we synthesized the regions.  There
are two isotopes of copper, $^{57}$Cu and $^{59}$Cu, both affected by
HFS. The HFS and IS information was taken from Hermann \etal{} (1993),
and the ratio of the isotopes seen in the solar system was used.
Unfortunately, the S/N in this
region is $\sim$ 25, even after co-adding the spectra from overlapping
orders. Using the formula in Fulbright \& Johnson (2003) to calculate the
expected EW errors, we find that we need to include a $\sigma=0.20$ dex
in log $\epsilon$ to account for the observational errors.

{\it Yttrium:} The \gfvalues{} are from Hannaford \etal{} (1982). Y has
one stable isotope, $^{89}$Y, and it has HFS. However, Johnson
\& Bolte (2001) investigated the effect of including HFS on the
measurement of the Y abundance in metal-poor stars, and found $<0.01$ dex
effect. Therefore, HFS has not been included in this study either.

{\it Zirconium:} The \gfvalues{} for this element come from
Bie\'mont \etal{} (1981).

{\it Barium:} We originally measured the abundance of Ba using
a set of strong lines ($\lambda$4554.0\AA, $\lambda$5853.7\AA, 
and $\lambda$6496.9\AA) that we have used in previous studies.
The \gfvalues{} and HFS information are from McWilliam (1998), and we
used the isotope ratios produced by the s-process from Arlandini \etal{} 
(1999).
The abundance from these three lines was log $\epsilon
= 2.61$, in disagreement with the results of A02 for
the same star (log $\epsilon=2.21$). This discrepancy cannot be explained
by the difference in model atmospheres. A02 used two different lines,
$\lambda$4130 and $\lambda$4166. When we used these two lines, we
derived a value of log$\epsilon=$2.46. This is an important point. With our high
Ba value, Eu/Ba agrees with the \sss. Our use of strong lines,
much more sensitive to $\xi$, could be part of the problem. NLTE effects
on the \ion{Ba}{2} lines could also be important. However, we cannot rule
out the possibility that the Ba abundance in this star is high. Our 
quoted error includes the contributions of EW errors and atmosphere
parameters errors, but not systematic NLTE effects.

{\it Lanthanum:} Lanthanum has only one stable isotope, but HFS is important.
We adopt both the \gfvalues{} and the HFS constants from the laboratory
study of Lawler, Bonvallet, \& Sneden (2001).

{\it Cerium:} \ion{Ce}{2} has only two isotopes, $^{140}$Ce and $^{142}$Ce, that
have a $>$1\% contribution to the solar-system abundance. 
While our \ion{Ce}{2} lines are weak, when we derived abundances
assuming lines without splitting, the stronger lines gave higher abundances,
which may indicate that IS cannot be ignored. We eliminated
all lines with EW$>$40m\AA. 
Doing this brought the average Ce abundance
down by 0.03 dex. We also tried including the approximate IS scheme of
Aoki \etal{} (2001), where the $^{142}$Ce line is shifted 0.11\AA{} relative
to the $^{140}$Ce line. We used the solar system ratios for the relative
abundances of the two Ce isotopes. This resulted in a change of 0.01 dex
in the average abundance. The \gfvalues{} for this element are from Palmeri
\etal{} (2000).

{\it Praseodymium:} New laboratory measurements of \gfvalues{} are
available for Scholl \etal{} (2002). 
HFS constants are available for all ten lines
we measured. The values of Li \etal{} (2000) or Rivest \etal{} (2002)
were used if measured; otherwise the values from Ginibre (1989) were
used.

{\it Neodymium: } Our first choice for \gfvalues{} for Nd were 
derived using lifetimes from Scholl \etal{} (2002) or Pinciuc \etal{}
(2001)  and branching
ratios from Maier \& Whaling (1977). Otherwise our \gfvalues{} were
adopted from KUR95. 
 There are seven stable Nd isotopes, and we used the solar system isotopic
abundances. While two of the
isotopes have odd Z, we neglect HFS for our lines since we could not
find HFS constants in the literature. We adopted IS from
Nakhate, Afzal, \& Ahmad (1997) if available or from Blaise 
\etal{} (1984). All but four of our
transitions were covered. The splitting ratios were taken from Aoki
\etal{} (2001).

{\it Samarium:} The \gfvalues{} are either taken directly from
Bi\'emont \etal{} (1989) or were computed using the branching
ratios of Saffman \& Whaling (1979) and the lifetimes of
Bi\'emont \etal{} (1989) or Vogel \etal{} (1988). We have IS information
for most lines. The IS
are from Villemoes \etal{} (1995) and Rao \etal{} (1990).  The splittings
ratios are given in Villemoes \etal{} (1995), and we used the solar
system isotope ratios. Two Sm isotopes, $^{147}$Sm and $^{149}$Sm have
HFS. We were unable to find HFS constants for our transitions, but
these two isotopes contribute $<30$\% to the solar system abundances.

{\it Europium:} The five lines we analyzed have \gfvalues, HFS constants,
IS constants in Lawler \etal{} (2001b). $^{151}$Eu and $^{153}$Eu are produced
in approximately equal proportions in the r-process and the s-process 
(Arlandini \etal{} 1999) so we have adopted a 50:50 split for our analysis.

{\it Gadolinium:} Our first choice for \gfvalues{} was Bergstrom
\etal{} (1988). We added additional oscillator strengths from the
compilation of Corliss \& Bozman (1962), following Bergstrom \etal's suggestion
that the log gf values of CB should be increased by 0.11 dex. We found that
the IS of the seven stable isotopes of Gd had an effect on the derived
abundances if the EW was too large. Since we could not find IS for all of
the transitions, we use only those lines with EW $<$ 35m\AA{} or with IS
information. This information was gathered from Brix \etal (1952), Ahmad, Saksena,
\& Venugopalan (1976), Ahmad, Venugopalan, \& Saksena (1979), and
Venugopalan, Afzal, \& Ahmad (1998). The isotopic splitting ratios are
from Kopfermann, Kruger, \& Steudel (1957).

{\it Terbium:} There is only one stable isotope of Tb, $^{159}$Tb. 
We adopted \gfvalues{} from Lawler \etal{} (2001a) and HFS constants from
Lawler, Wyart, \& Blaise (2001). 

{\it Dysprosium:}
There are seven stable isotopes of Dy, and all are made
in the s-process. Two of the isotopes have odd Z and so have hyperfine
splitting, but we couldn't find information on our transitions. We did
find information on IS for most of our transitions in Aufmuth (1978). We
used the solar system isotope ratios.
The \gfvalues{} are from Wickliffe, Lawler and Nave (2000)

{\it Holmium: } There is only one stable isotope of Ho, $^{165}$Ho, but
HFS is important. For $\lambda$4045, we use the Sneden \etal{} (1996) linelist,
with their HFS and \gfvalue.  For $\lambda$3398 and $\lambda$3416, we
adopt HFS constants from Worm, Shi, \& Poulsen (1990). The \gfvalue{}
of $\lambda$3398 comes from Gorshkov \& Komarovskii (1979), while the
\gfvalue of $\lambda$3416 was taken from the Kurucz linelist.

{\it Erbium:} We confine our analysis to lines with oscillator strengths 
from Musiol \& Labuz (1983). We were unable to
find IS or HFS information for 
the observed transitions in the literature.

{\it Thulium:} Wickliffe \& Lawler (1997) have recently published high-quality
\gfvalues{} for Tm transitions, including the
four included in this study. No HFS information was available,
but all lines had EW $<$ 20 m\AA, so our abundance analysis should
be accurate.

{\it Ytterbium:} The \gfvalue{} for the one line of Yb we could use is
from Pinnington, Rieger, \& Kernahan (1997). The IS and HFS information is
from M\aa rtensson-Pendrill, Gough, \& Hannaford (1994). The isotope
ratios are the predicted s-process ones from Arlandini \etal{} (1999).

{\it Hafnium:} There are five isotopes of Hf; two of those have
hyperfine splitting. We could only find information for the strongest line
(3399.8\AA) out of the eleven we used (Zhao \etal{} 1997).
The HFS is considerably larger than the isotopic
splitting. The use of the IS (with solar system ratios)
and HFS reduces the abundance derived from the 3399.8\AA{} by
0.64 dex. While the other lines have no information, they are weaker 
(EW $<$ 35m\AA), and their average abundance is equal to that derived
from the 3399\AA. 

{\it Lead:} We adopt the line lists from Johnson \& Bolte (2002a).

\include{tab1}

\end{document}

%% file: tab1.tex
\pagestyle{empty}
\singlespace
\begin{deluxetable}{lrrcrrr}
\tablenum{1}
\tablewidth{0pt}
\tablecaption{Equivalent Widths}
\tablehead{
\colhead{$\lambda$} & \colhead {E.P.} & \colhead {log $gf$} & \colhead{Method} &\colhead {EW} & \colhead {HFS,} & \colhead {log $\epsilon$} \\
\colhead {\AA} & \colhead{(eV)}& \colhead{} & \colhead{} & \colhead {m\AA} & 
\colhead{IS?} & \colhead{} 
}
\startdata
\multicolumn{7}{c}{MgI} \\
   4703.00 &    4.34 &   $-$0.523 & EW &    73.8  &   &    6.01 \\
\multicolumn{7}{c}{CaI} \\
   4318.65 &    1.90 &   $-$0.208 & EW &    43.5  &   &    4.36 \\
   5261.71 &    2.52 &   $-$0.580 & EW &    12.7  &   &    4.50 \\
\multicolumn{7}{c}{TiI} \\
   3653.50 &    0.05 &    0.280 & EW &    34.8  &   &    2.52 \\
   3729.81 &    0.00 &   $-$0.290 & EW &    38.0  &   &    3.10 \\
   3741.06 &    0.02 &   $-$0.150 & EW &    41.1  &   &    3.05 \\
   3752.86 &    0.05 &    0.040 & EW &    23.1  &   &    2.45 \\
   4533.24 &    0.85 &    0.540 & EW &    23.8  &   &    2.71 \\
   4534.78 &    0.84 &    0.340 & EW &    22.1  &   &    2.86 \\
   4617.27 &    1.75 &    0.450 & EW &     7.5  &   &    3.11 \\
   4981.73 &    0.85 &    0.560 & EW &    39.4  &   &    3.02 \\
   5210.39 &    0.05 &   $-$0.820 & EW &    13.1  &   &    2.85 \\
\multicolumn{7}{c}{TiII} \\
   4450.48 &    1.08 &   $-$1.510 & EW &    39.5  &   &    2.72 \\
   4468.49 &    1.13 &   $-$0.600 & EW &    80.6  &   &    2.91 \\
   4501.27 &    1.12 &   $-$0.760 & EW &    74.6  &   &    2.89 \\
   4657.20 &    1.24 &   $-$2.320 & EW &    12.1  &   &    2.94 \\
\multicolumn{7}{c}{VII} \\
   3545.19 &    1.10 &   $-$0.259 & EW &    26.5  &   &    1.30 \\
   3592.02 &    1.10 &   $-$0.263 & EW &    37.5  &   &    1.58 \\
\multicolumn{7}{c}{CrI} \\
   4600.76 &    1.00 &   $-$1.260 & EW &    13.4  &   &    3.57 \\
   4626.18 &    0.97 &   $-$1.320 & EW &     6.0  &   &    3.20 \\
   4646.17 &    1.03 &   $-$0.720 & EW &    21.2  &   &    3.31 \\
\multicolumn{7}{c}{MnI} \\
   3823.51 &    2.14 &   $-$0.513 & EW &    13.8  & HFS  &    2.86 \\
   4030.74 &    0.00 &   $-$1.037 & EW &    89.0  & HFS  &    2.64 \\
   4033.09 &    0.00 &   $-$1.291 & EW &    77.0  & HFS  &    2.57 \\
   4754.04 &    2.28 &   $-$0.677 & EW &     6.9  & HFS  &    2.69 \\
\multicolumn{7}{c}{FeI} \\
   3617.78 &    3.02 &   $-$0.050 & EW  &   43.3  &   &    5.07 \\
   3709.25 &    0.92 &   $-$0.610 & EW  &   99.0  &   &    5.04 \\
   3715.91 &    2.28 &   $-$1.530 & EW  &   20.9  &   &    5.08 \\
   3760.05 &    2.40 &   $-$0.810 & EW  &   38.2  &   &    4.93 \\
   3767.19 &    1.01 &   $-$0.350 & EW  &   96.4  &   &    4.79 \\
   3787.88 &    1.01 &   $-$0.820 & EW  &   86.4  &   &    4.97 \\
   3790.10 &    0.99 &   $-$1.720 & EW  &   56.3  &   &    4.86 \\
   3856.37 &    0.05 &   $-$1.250 & EW  &  116.0  &   &    5.11 \\
   3899.71 &    0.09 &   $-$1.490 & EW  &   95.4  &   &    4.92 \\
   3906.48 &    0.11 &   $-$2.200 & EW  &   81.5  &   &    5.22 \\
   3916.74 &    3.24 &   $-$0.560 & EW  &   20.9  &   &    5.06 \\
   3917.18 &    0.99 &   $-$2.150 & EW  &   55.0  &   &    5.22 \\
   3922.91 &    0.05 &   $-$1.610 & EW  &   97.6  &   &    5.05 \\
   3949.96 &    2.17 &   $-$1.250 & EW  &   34.7  &   &    5.02 \\
   4114.45 &    2.83 &   $-$1.300 & EW  &   13.0  &   &    5.10 \\
   4147.68 &    1.49 &   $-$2.060 & EW  &   41.7  &   &    5.25 \\
   4156.81 &    2.83 &   $-$0.810 & EW  &   38.7  &   &    5.32 \\
   4187.04 &    2.45 &   $-$0.510 & EW  &   57.3  &   &    5.10 \\
   4202.03 &    1.49 &   $-$0.670 & EW  &   87.3  &   &    5.19 \\
   4216.19 &    0.00 &   $-$3.320 & EW  &   43.8  &   &    4.99 \\
   4222.22 &    2.45 &   $-$0.930 & EW  &   38.2  &   &    5.03 \\
   4233.61 &    2.48 &   $-$0.560 & EW  &   51.3  &   &    5.01 \\
   4250.13 &    2.47 &   $-$0.370 & EW  &   67.2  &   &    5.25 \\
   4494.57 &    2.20 &   $-$1.100 & EW  &   46.6  &   &    5.11 \\
   4531.15 &    1.49 &   $-$2.110 & EW  &   36.7  &   &    5.14 \\
   4871.32 &    2.85 &   $-$0.360 & EW  &   53.3  &   &    5.17 \\
   4872.14 &    2.87 &   $-$0.570 & EW  &   39.8  &   &    5.08 \\
   4891.49 &    2.84 &   $-$0.110 & EW  &   67.7  &   &    5.29 \\
   4903.31 &    2.87 &   $-$0.930 & EW  &   23.9  &   &    5.06 \\
   4920.50 &    2.82 &    0.070 & EW  &   66.0  &   &    5.04 \\
   4994.13 &    0.92 &   $-$3.040 & EW  &   25.1  &   &    5.15 \\
   5006.12 &    2.83 &   $-$0.660 & EW  &   41.9  &   &    5.17 \\
   5041.07 &    0.95 &   $-$3.090 & EW  &   17.5  &   &    5.02 \\
   5049.82 &    2.28 &   $-$1.340 & EW  &   44.0  &   &    5.32 \\
   5051.64 &    0.92 &   $-$2.760 & EW  &   31.7  &   &    5.03 \\
   5166.29 &    0.00 &   $-$4.160 & EW  &   16.7  &   &    5.06 \\
   5171.60 &    1.49 &   $-$1.750 & EW  &   54.8  &   &    5.14 \\
   5192.34 &    2.99 &   $-$0.420 & EW  &   43.3  &   &    5.11 \\
   5194.94 &    1.56 &   $-$2.050 & EW  &   36.9  &   &    5.10 \\
   5198.71 &    2.22 &   $-$2.090 & EW  &    9.2  &   &    5.01 \\
\enddata
\end{deluxetable}

\begin{deluxetable}{lrrcrrr}
\tablenum{1}
\tablewidth{0pt}
\tablecaption{Equivalent Widths}
\tablehead{
\colhead{$\lambda$} & \colhead {E.P.} & \colhead {log $gf$} & \colhead{Method} &\colhead {EW} & \colhead {HFS,} & \colhead {log $\epsilon$} \\
\colhead {\AA} & \colhead{(eV)}& \colhead{} & \colhead{} & \colhead {m\AA} & 
\colhead{IS?} & \colhead{} 
}
\startdata
   5216.28 &    1.61 &   $-$2.110 & EW  &   25.9  &   &    4.95 \\
   5217.39 &    3.21 &   $-$1.070 & EW  &   12.9  &   &    5.18 \\
   5232.94 &    2.94 &   $-$0.100 & EW  &   57.3  &   &    5.07 \\
   5242.49 &    3.62 &   $-$0.970 & EW  &    5.0  &   &    5.04 \\
   5266.56 &    2.99 &   $-$0.380 & EW  &   37.1  &   &    4.93 \\
   5269.54 &    0.86 &   $-$1.330 & EW  &   94.6  &   &    5.18 \\
\multicolumn{7}{c}{FeII} \\
   4491.40 &    2.86 &   $-$2.710 & EW   &   14.2  &   &    5.03 \\
   4508.29 &    2.86 &   $-$2.330 & EW  &   26.0  &   &    5.00 \\
   4520.23 &    2.81 &   $-$2.600 & EW  &   26.9  &   &    5.24 \\
   4555.89 &    2.83 &   $-$2.390 & EW  &   21.3  &   &    4.90 \\
   4582.84 &    2.84 &   $-$3.100 & EW  &   10.1  &   &    5.21 \\
   4583.84 &    2.81 &   $-$1.920 & EW  &   49.4  &   &    5.08 \\
   4629.34 &    2.81 &   $-$2.370 & EW  &   31.6  &   &    5.12 \\
   5018.45 &    2.89 &   $-$1.220 & EW  &   80.5  &   &    5.27 \\
   5197.56 &    3.23 &   $-$2.100 & EW  &   19.8  &   &    4.94 \\
   5234.62 &    3.22 &   $-$2.230 & EW  &   24.7  &   &    5.19 \\
\multicolumn{7}{c}{CoI} \\
   4121.30 &    0.92 &   $-$0.993 & EW &    37.4  & HFS  &    2.60 \\
\multicolumn{7}{c}{NiI} \\
   3500.85 &    0.17 &   $-$1.294 & EW &    56.8  &   &    3.50 \\
   3515.05 &    0.11 &   $-$0.266 & EW &    94.6  &   &    3.67 \\
   3519.77 &    0.28 &   $-$1.422 & EW &    53.6  &   &    3.62 \\
   3524.54 &    0.03 &   $-$0.007 & EW &   105.5  &   &    3.55 \\
   3571.87 &    0.17 &   $-$1.154 & EW &    76.5  &   &    4.05 \\
   3597.71 &    0.21 &   $-$1.115 & EW &    66.0  &   &    3.67 \\
   3602.28 &    0.17 &   $-$2.192 & EW &    31.3  &   &    3.60 \\
   3612.74 &    0.28 &   $-$1.423 & EW &    58.7  &   &    3.77 \\
   3619.39 &    0.42 &    0.020 & EW &    87.9  &   &    3.49 \\
   3664.10 &    0.28 &   $-$2.130 & EW &    35.3  &   &    3.67 \\
   3858.30 &    0.42 &   $-$0.951 & EW &    77.2  &   &    3.87 \\
\multicolumn{7}{c}{CuI} \\
   3247.58 &    0.00 &   $-$0.421 & syn  &   58.1  & HFS,IS  &    0.59 \\
   3273.98 &    0.00 &   $-$0.864 & syn  &   54.0  & HFS,IS  &    0.79 \\
\multicolumn{7}{c}{ZnI} \\
   4810.55 &    4.08 &   $-$0.170 & EW  &    8.8  &   &    2.32 \\
\multicolumn{7}{c}{YII} \\
   3549.01 &    0.13 &   $-$0.280 & EW  &   38.8  &   &    0.23 \\
   3584.52 &    0.10 &   $-$0.410 & syn  &   39.4  &   &    0.34 \\
   3600.74 &    0.18 &    0.280 & syn  &   53.2  &   &    0.19 \\
   3601.91 &    0.10 &   $-$0.180 & syn  &   45.5  &   &    0.29 \\
   3611.04 &    0.13 &    0.010 & syn  &   45.9  &   &    0.14 \\
   3628.70 &    0.13 &   $-$0.710 & syn  &   26.3  &   &    0.29 \\
   3710.29 &    0.18 &    0.460 & syn  &   64.9   &   &    0.29 \\
   3774.34 &    0.13 &    0.210 & syn  &   57.5   &   &    0.19 \\
   3950.36 &    0.10 &   $-$0.490 & syn  &   38.5   &   &    0.24 \\
   4854.87 &    0.99 &   $-$0.380 & EW  &   18.1   &   &    0.41 \\
   4883.69 &    1.08 &    0.070 & EW  &   32.2   &   &    0.43 \\
   4900.11 &    1.03 &   $-$0.090 & syn  &   28.0   &   &    0.44 \\
   5087.42 &    1.08 &   $-$0.170 & syn  &   23.5   &   &    0.44 \\
   5200.41 &    0.99 &   $-$0.570 & EW  &   10.1   &   &    0.26 \\
\multicolumn{7}{c}{ZrII} \\
   3458.94 &    0.96 &   $-$0.520 & syn  &   10.3   &   &    0.91 \\
   3479.39 &    0.71 &    0.170 & syn  &   36.0   &   &    0.81 \\
   3505.67 &    0.16 &   $-$0.360 & EW  &   44.1   &   &    1.00 \\
   3551.96 &    0.09 &   $-$0.310 & syn  &   51.3   &   &    1.11 \\
   4050.33 &    0.71 &   $-$1.000 & EW  &    9.1   &   &    0.93 \\
   4208.98 &    0.71 &   $-$0.460 & EW  &   35.0   &   &    1.19 \\
   4258.05 &    0.56 &   $-$1.130 & syn  &   15.7   &   &    1.16 \\
   4443.00 &    1.49 &   $-$0.330 & syn  &   10.6   &   &    1.11 \\
   4496.97 &    0.71 &   $-$0.810 & syn  &   23.4   &   &    1.21 \\
\multicolumn{7}{c}{PdI} \\
   3404.58 &    0.81 &    0.320 & syn  &   9.2   &   &    0.17 \\
\multicolumn{7}{c}{LaII} \\
   3628.82 &    0.13 &   $-$1.763 & syn  &   21.5   & HFS  &    0.84 \\
   3713.57 &    0.17 &   $-$1.372 & syn  &   40.2   & HFS  &    0.74 \\
   3794.77 &    0.24 &   $-$0.473 & syn  &   56.5   & HFS  &    0.60 \\
   3929.24 &    0.17 &   $-$0.962 & syn  &   81.6   & HFS  &    1.04 \\
   3988.56 &    0.40 &   $-$0.475 & EW  &  129.1   & HFS  &    0.89 \\
   3995.78 &    0.17 &   $-$0.706 & EW  &   94.1   & HFS  &    0.96 \\
   4086.71 &    0.00 &   $-$0.696 & syn  &   71.6   & HFS  &    0.90 \\
   4123.22 &    0.32 &   $-$0.482 & syn  &   75.4   & HFS  &    0.95 \\
\enddata
\end{deluxetable}

\begin{deluxetable}{lrrcrrr}
\tablenum{1}
\tablewidth{0pt}
\tablecaption{Equivalent Widths}
\tablehead{
\colhead{$\lambda$} & \colhead {E.P.} & \colhead {log $gf$} & \colhead{Method} &\colhead {EW} & \colhead {HFS,} & \colhead {log $\epsilon$} \\
\colhead {\AA} & \colhead{(eV)}& \colhead{} & \colhead{} & \colhead {m\AA} & 
\colhead{IS?} & \colhead{} 
}
\startdata
   4322.54 &    0.17 &   $-$1.532 & syn  &   52.8   & HFS  &    0.99 \\
   4526.10 &    0.77 &   $-$1.222 & EW  &   25.8   & HFS  &    0.92 \\
   4558.46 &    0.32 &   $-$1.572 & syn  &   27.3   & HFS  &    0.84 \\
   4574.95 &    0.17 &   $-$1.706 & EW  &   39.6   & HFS  &    0.88 \\
   4580.04 &    0.71 &   $-$1.390 & syn  &   12.3   & HFS  &    0.84 \\
   4613.37 &    0.71 &   $-$1.283 & syn  &   29.0   & HFS  &    1.09 \\
   4645.31 &    0.13 &   $-$2.455 & syn  &    9.0   & HFS  &    0.84 \\
   4662.52 &    0.00 &   $-$1.763 & syn  &   32.7   & HFS  &    0.89 \\
   4716.41 &    0.77 &   $-$1.733 & EW  &    8.3   & HFS  &    0.87 \\
   4728.45 &    0.17 &   $-$1.982 & syn  &   28.0   & HFS  &    0.99 \\
   4740.28 &    0.13 &   $-$1.652 & EW  &   40.6   & HFS  &    1.07 \\
   4804.01 &    0.23 &   $-$2.084 & EW  &   28.4   & HFS  &    1.17 \\
   4808.99 &    0.23 &   $-$1.780 & EW  &   20.6   & HFS  &    0.92 \\
   4824.03 &    0.65 &   $-$1.250 & syn  &   25.2   & HFS  &    0.99 \\
   4840.01 &    0.32 &   $-$2.416 & EW  &   10.3   & HFS  &    1.10 \\
   4970.39 &    0.32 &   $-$1.683 & syn  &   26.0   & HFS  &    0.99 \\
   4986.86 &    0.17 &   $-$1.823 & syn  &   25.7   & HFS  &    0.84 \\
\multicolumn{7}{c}{CeII} \\
   3534.06 &    0.52 &   $-$0.210 &   &   23.0  &   &    1.20 \\
   3539.08 &    0.32 &   $-$0.500 &   &   28.0  &   &    1.43 \\
   3577.47 &    0.47 &    0.080 &   &   34.8  &   &    1.21 \\
   3653.66 &    0.47 &   $-$0.190 &   &   19.3  &   &    0.93 \\
   3655.85 &    0.32 &   $-$0.150 &   &   29.5  &   &    1.04 \\
   3659.23 &    0.17 &   $-$0.790 &   &   21.4  &   &    1.28 \\
   3709.95 &    0.12 &   $-$0.320 &   &   36.6  &   &    1.19 \\
   3764.11 &    0.36 &   $-$0.220 &   &   35.0  &   &    1.28 \\
   3838.54 &    0.33 &   $-$0.100 &   &   42.5  &   &    1.33 \\
   3896.78 &    0.56 &   $-$0.410 &   &   18.5  &   &    1.18 \\
   3919.82 &    0.70 &   $-$0.220 &   &   26.7  &   &    1.38 \\
   3923.11 &    0.56 &   $-$0.630 &   &   24.5  &   &    1.58 \\
   3924.65 &    0.56 &   $-$0.280 &   &   19.5  &   &    1.08 \\
   3940.36 &    0.32 &   $-$0.380 &   &   33.9  &   &    1.34 \\
   3942.75 &    0.86 &    0.670 &   &   40.3  &   &    1.03 \\
   3960.92 &    0.32 &   $-$0.460 &   &   26.2  &   &    1.20 \\
   4003.77 &    0.93 &    0.200 &   &   24.6  &   &    1.13 \\
   4031.34 &    0.32 &   $-$0.150 &   &   33.0  &   &    1.06 \\
   4042.59 &    0.50 &    0.110 &   &   41.7  &   &    1.23 \\
   4046.34 &    0.55 &   $-$0.710 &   &   26.6  &   &    1.69 \\
   4053.49 &    0.00 &   $-$0.800 &   &   29.8  &   &    1.28 \\
   4062.23 &    1.37 &    0.300 &   &   16.7  &   &    1.24 \\
   4068.83 &    0.70 &   $-$0.220 &   &   35.7  &   &    1.60 \\
   4073.49 &    0.48 &    0.180 &   &   44.4  &   &    1.21 \\
   4075.71 &    0.70 &    0.270 &   &   36.5  &   &    1.13 \\
   4083.23 &    0.70 &    0.190 &   &   39.1  &   &    1.29 \\
   4115.38 &    0.92 &    0.050 &   &   24.6  &   &    1.26 \\
   4118.15 &    0.70 &    0.140 &   &   37.5  &   &    1.29 \\
   4127.38 &    0.68 &    0.300 &   &   42.3  &   &    1.23 \\
   4137.65 &    0.52 &    0.390 &   &   55.1  &   &    1.38 \\
   4145.00 &    0.70 &    0.090 &   &   33.9  &   &    1.23 \\
   4148.92 &    1.09 &   $-$0.010 &   &   13.7  &   &    1.14 \\
   4165.60 &    0.91 &    0.480 &   &   42.0  &   &    1.28 \\
   4444.40 &    0.92 &   $-$0.110 &   &   21.4  &   &    1.29 \\
   4444.70 &    1.06 &    0.080 &   &   24.6  &   &    1.33 \\
   4460.23 &    0.48 &    0.270 &   &   58.0  &   &    1.48 \\
   4486.91 &    0.30 &   $-$0.330 &   &   41.0  &   &    1.36 \\
   4523.08 &    0.52 &   $-$0.080 &   &   50.4  &   &    1.61 \\
   4528.48 &    0.86 &    0.330 &   &   34.8  &   &    1.13 \\
   4539.78 &    0.33 &   $-$0.080 &   &   43.1  &   &    1.19 \\
   4560.97 &    0.68 &   $-$0.220 &   &   20.1  &   &    1.09 \\
   4565.86 &    1.09 &   $-$0.010 &   &   19.3  &   &    1.29 \\
   4593.94 &    0.70 &    0.070 &   &   49.8  &   &    1.63 \\
   4562.37 &    0.48 &    0.190 &   &   52.9  &   &    1.37 \\
   4628.16 &    0.52 &    0.150 &   &   47.6  &   &    1.28 \\
   4773.96 &    0.92 &   $-$0.330 &   &   11.7  &   &    1.13 \\
   4882.49 &    1.53 &    0.200 &   &   11.3  &   &    1.21 \\
   5044.03 &    1.21 &   $-$0.140 &   &    8.5  &   &    1.06 \\
   5187.46 &    1.21 &    0.080 &   &   19.3  &   &    1.27 \\
   5274.24 &    1.04 &    0.080 &   &   23.8  &   &    1.21 \\
\enddata
\end{deluxetable}

\begin{deluxetable}{lrrcrrr}
\tablenum{1}
\tablewidth{0pt}
\tablecaption{Equivalent Widths}
\tablehead{
\colhead{$\lambda$} & \colhead {E.P.} & \colhead {log $gf$} & \colhead{Method} &\colhead {EW} & \colhead {HFS,} & \colhead {log $\epsilon$} \\
\colhead {\AA} & \colhead{(eV)}& \colhead{} & \colhead{} & \colhead {m\AA} & 
\colhead{IS?} & \colhead{} 
}
\startdata
\multicolumn{7}{c}{PrII} \\
   3964.88 &    0.05 &   $-$0.503 &  SYN &   41.2   & HFS  &    0.04 \\
   3965.33 &    0.20 &   $-$0.432 &  SYN &   41.0   & HFS  &    0.11 \\
   4044.75 &    0.00 &   $-$0.803 &  SYN &   21.7   & HFS  &   $-$0.01 \\
   4062.74 &    0.42 &   $-$0.148 &  SYN &   43.6   & HFS  &    0.09 \\
   4143.20 &    0.37 &   $-$0.098 &  SYN &   67.0   & HFS  &    0.19 \\
   4179.47 &    0.20 &   $-$0.248 &  SYN &   74.9   & HFS  &    0.24 \\
   5173.89 &    0.97 &   $-$0.179 &  SYN &   12.1   & HFS  &   $-$0.01 \\
   5219.03 &    0.80 &   $-$0.812 &  SYN &    8.9   & HFS  &    0.29 \\
   5220.10 &    0.80 &   $-$0.417 &  SYN &   17.1   & HFS  &    0.24 \\
   5259.72 &    0.63 &   $-$0.502 &  SYN &   16.6   & HFS  &    0.14 \\
\multicolumn{7}{c}{NdII} \\
   3713.70 &    0.74 &    0.080 & SYN  &   21.5   &   &    0.75 \\
   3715.69 &    0.47 &   $-$0.190 & SYN  &   27.2   & IS  &    0.90 \\
   3780.40 &    0.47 &   $-$0.130 & SYN  &   27.9   & IS  &    0.84 \\
   3784.25 &    0.38 &    0.390 &  EW &   50.7   & IS  &    0.92 \\
   3900.23 &    0.47 &    0.340 & SYN  &   45.2   & IS  &    0.86 \\
   3941.51 &    0.06 &   $-$0.010 & SYN  &   51.8   & IS  &    0.96 \\
   3957.99 &    0.06 &   $-$0.650 & SYN  &   35.2   & IS  &    1.09 \\
   3973.25 &    0.63 &    0.210 & EW  &   47.9   & IS  &    1.19 \\
   3976.12 &    0.21 &   $-$1.200 & SYN  &   13.4   & IS  &    1.14 \\
   3976.84 &    0.00 &   $-$0.590 & SYN  &   43.2   & IS  &    1.18 \\
   3979.47 &    0.21 &   $-$0.280 & SYN  &   49.5   & IS  &    1.25 \\
   3992.59 &    1.41 &    0.160 & SYN  &   17.5   &   &    1.20 \\
   4011.08 &    0.47 &   $-$0.830 &  EW &   12.4   &   &    1.00 \\
   4012.23 &    0.63 &    0.780 & SYN  &   74.1   & IS  &    1.14 \\
   4012.70 &    0.00 &   $-$0.740 & SYN  &   36.5   & IS  &    1.15 \\
   4013.23 &    0.18 &   $-$1.150 &  EW &   14.8   &   &    1.11 \\
   4018.82 &    0.06 &   $-$0.890 & SYN  &   28.6   & IS  &    1.15 \\
   4020.86 &    0.32 &   $-$0.270 & SYN  &   45.3   &IS   &    1.23 \\
   4021.33 &    0.32 &    0.230 & SYN  &   44.2   & IS  &    0.75 \\
   4021.75 &    0.18 &   $-$0.300 & SYN  &   42.4   & IS  &    1.05 \\
   4023.00 &    0.21 &   $-$0.200 & SYN  &   47.0   & IS  &    1.14 \\
   4043.59 &    0.32 &   $-$0.510 &  EW &   24.3   & IS  &    0.92 \\
   4051.14 &    0.38 &    0.090 &  EW &   39.5   & IS  &    0.80 \\
   4061.07 &    0.47 &    0.520 & SYN  &   67.8   & IS  &    1.14 \\
   4109.44 &    0.00 &   $-$0.810 & SYN  &   25.1   & IS  &    0.86 \\
   4113.83 &    0.18 &   $-$0.900 & SYN  &   29.6   & IS  &    1.29 \\
   4133.35 &    0.32 &   $-$0.510 & SYN  &   32.8   & IS  &    1.12 \\
   4156.07 &    0.18 &    0.180 & EW  &   69.9   & IS  &    1.17 \\
   4211.29 &    0.21 &   $-$0.720 & SYN  &   30.0   & IS  &    1.15 \\
   4446.37 &    0.21 &   $-$0.590 & EW  &   46.5   & IS  &    1.38 \\
   4451.55 &    0.38 &    0.110 & SYN  &   61.4   & IS  &    1.13 \\
   4451.98 &    0.00 &   $-$1.340 &  EW &   26.6   & IS  &    1.41 \\
   4567.60 &    0.21 &   $-$1.510 & EW  &    9.3   & IS  &    1.19 \\
   4579.32 &    0.74 &   $-$0.650 & SYN  &   15.0   & IS  &    1.15 \\
   4645.77 &    0.56 &   $-$0.750 & EW  &   20.4   & IS  &    1.22 \\
   5092.78 &    0.38 &   $-$0.610 & EW  &   28.1   & IS  &    1.04 \\
   5192.62 &    1.14 &    0.310 & SYN  &   35.2   & IS   &    1.10 \\
   5200.12 &    0.56 &   $-$0.490 &  EW &   10.7   & IS  &    0.56 \\
   5212.35 &    0.21 &   $-$0.870 &  EW &   24.2   & IS  &    1.02 \\
\multicolumn{7}{c}{SmII} \\
   3568.28  &    0.48  &   0.290  & SYN &  36.9 & IS  &  0.42 \\
   3609.49  &    0.28  &   0.140  & SYN &  34.3 & IS &  0.41 \\
   3634.29  &    0.18  &   0.020  & SYN &  44.9 &  &  0.80 \\
   3706.75  &    0.48  &  $-$0.630  & SYN  &  8.6 &  &  0.45 \\
   3718.88  &    0.38  &  $-$0.350  & EW  &  22.8 &  IS &  0.60 \\
   3896.97  &    0.04  &  $-$0.580  & SYN &  20.9  & &  0.40 \\
   4244.70  &    0.28  &  $-$0.730  & SYN  &  11.3  &IS &  0.40 \\
   4499.48  &    0.25  &  $-$1.010  & EW &  12.6  & IS &  0.68 \\
   4519.63  &    0.54  &  $-$0.430  & SYN &  19.5  & IS &  0.61 \\
   4523.90  &    0.43  &  $-$0.580 & EW  &  21.1  & IS &  0.70 \\
   4577.69  &    0.25  &  $-$0.770  & EW  &  16.6 & IS  &  0.58 \\
   4815.81  &    0.18  &  $-$0.760 & EW  &  13.9  & &  0.37 \\
\multicolumn{7}{c}{EuII} \\
   3907.09 &    0.21 &   $-$0.374 & SYN   &   72.0   & HFS,IS  &   $-$0.10 \\
   3930.50 &    0.21 &   $-$0.326 & SYN  &   87.8   & HFS,IS  &   $-$0.10 \\
   4129.76 &    0.00 &   $-$0.401 & SYN  &  115.0   & HFS,IS  &   $-$0.12 \\
   4205.04 &    0.00 &   $-$0.386 & SYN  &  106.3   & HFS,IS  &   $-$0.10 \\
   4435.53 &    0.21 &   $-$0.696 & SYN  &   78.2   & HFS,IS  &    0.05 \\
\enddata
\end{deluxetable}

\begin{deluxetable}{lrrcrrr}
\tablenum{1}
\tablewidth{0pt}
\tablecaption{Equivalent Widths}
\tablehead{
\colhead{$\lambda$} & \colhead {E.P.} & \colhead {log $gf$} & \colhead{Method} &\colhead {EW} & \colhead {HFS,} & \colhead {log $\epsilon$} \\
\colhead {\AA} & \colhead{(eV)}& \colhead{} & \colhead{} & \colhead {m\AA} & 
\colhead{IS?} & \colhead{} 
}
\startdata
\multicolumn{7}{c}{GdII} \\
   3331.40  &    0.00  &  $-$0.140 &  SYN  & 28.6  &  &  0.47 \\
   3392.50  &    0.08  &  $-$0.220 &  SYN  & 25.0  &  &  0.51 \\
   3418.70  &    0.00  &  $-$0.310 &  EW  & 23.2  &  &  0.45 \\
   3423.92  &    0.00  &  $-$0.410 &  SYN  & 20.6  &   & 0.46 \\
   3424.59  &    0.35  &  $-$0.170  & SYN  & 19.3  & IS & 0.54 \\
   3439.21  &    0.38  &   0.150  &  EW & 24.7  & IS  &   0.42 \\
   3451.24  &    0.38  &  $-$0.050  &  SYN & 18.2  &  & 0.42 \\
   3454.91  &    0.03  &  $-$0.590  & SYN  & 18.8  &  &  0.61 \\
   3464.00  &    0.43  &   0.390  & SYN  & 37.9  & IS  &  0.54 \\
   3467.27  &    0.43  &   0.150  & SYN  & 28.4  & IS  &  0.58 \\
   3468.99  &    0.43  &   0.150  & SYN  & 30.7  &  &  0.66 \\
   3473.22  &    0.03  &  $-$0.300  & EW  & 32.4  &  &  0.74 \\
   3481.28  &    0.60  &   0.450  & SYN  & 30.0  &  &  0.52 \\
   3481.80  &    0.49  &   0.230  & SYN  & 24.7  & IS  &    0.45 \\
   3549.40  &    0.24  &   0.310  & EW  & 37.3  & IS  &   0.45 \\
   3654.62  &    0.08  &   0.070  & SYN  & 32.0  &  &   0.31 \\
   3699.74  &    0.35  &  $-$0.150  & SYN  & 18.7  &  &   0.41 \\
   3719.53  &    0.49  &   0.120  & SYN  & 15.4  & IS  & 0.17 \\
   3916.51  &    0.60  &   0.170  & SYN  & 28.6  &  &    0.61 \\
   4037.97  &    0.56  &  $-$0.230  & SYN  & 17.4  & IS  &  0.58 \\
   4063.38  &    0.99  &   0.470  & SYN &  24.0  & IS  &   0.55 \\
   4085.56  &    0.73  &   0.070  & EW &  19.9  & IS  &    0.57 \\
   4098.60  &    0.60  &   0.390  & SYN &  37.2  & IS  &  0.60 \\
   4130.40  &    0.73  &   0.210  & SYN &  29.7  & IS  &   0.68 \\
\multicolumn{7}{c}{TbII}\\   
   3509.18 &    0.00 &    0.162 & syn  &   29.60  & HFS  &   $-$0.70 \\
   3568.55 &    0.00 &   $-$0.179 & syn  &   12.60  & HFS  &   $-$0.80 \\
   3702.81 &    0.13 &   $-$0.091 & syn  &   14.20  & HFS  &   $-$0.75 \\
   4005.43 &    0.13 &   $-$0.551 & syn  &    5.10  & HFS  &   $-$0.80 \\
\multicolumn{7}{c}{DyII}\\
   3407.80  &    0.00  &   0.180  & syn &   44.0   & IS & 0.39 \\
   3434.37  &    0.00  &  $-$0.450  & EW  &   19.7   & IS & 0.21 \\
   3445.57  &    0.00  &  $-$0.150  & EW  &   37.4   & IS & 0.46 \\
   3454.32  &    0.10  &  $-$0.140  & syn &   32.1   &    & 0.40 \\
   3460.97  &    0.00  &  $-$0.070  & EW  &   35.8   & IS & 0.33 \\
   3523.98  &    0.54  &   0.420  & syn &   32.3   & IS & 0.28 \\
   3531.71  &    0.00  &   0.770  & syn &   60.4   & IS & 0.38 \\
   3534.96  &    0.10  &  $-$0.040  & syn &   32.7   & IS & 0.29 \\
   3536.02  &    0.54  &   0.530  & EW  &   38.6   & IS & 0.37 \\
   3538.52  &    0.00  &  $-$0.020  & EW  &   44.4   & IS & 0.53 \\
   3546.83  &    0.10  &  $-$0.550  & syn &   20.9   & IS & 0.44 \\
   3550.22  &    0.59  &   0.270  & EW  &   30.6   & IS & 0.43 \\
   3563.15  &    0.10  &  $-$0.360  & syn &   25.7   &    & 0.40 \\
   3694.81  &    0.10  &  $-$0.110  & syn &   37.3   & IS & 0.39 \\
   3747.82  &    0.10  &  $-$0.810  & syn &   16.1   &    & 0.45 \\
   3757.37  &    0.10  &  $-$0.170  & syn &   35.8   &    & 0.40 \\
   3841.31  &    0.10  &  $-$0.780  & syn &   13.4   &    & 0.30 \\
   3983.65  &    0.54  &  $-$0.310  & syn &   16.7   &    & 0.40 \\
   3996.69  &    0.59  &  $-$0.260  & syn &   18.1   &    & 0.45 \\
   4077.97  &    0.10  &  $-$0.040  & syn &   42.4   &    & 0.40 \\
\multicolumn{7}{c}{HoII}\\
   3399.03 &    0.00 &   $-$0.496 &   &   62.5   &   &   $-$0.22 \\
   3416.37 &    0.08 &   $-$0.523 &   &   26.7   &   &   $-$0.52 \\
   4045.49 &    0.00 &   $-$0.933 &   &   27.3   &   &   $-$0.32 \\
\multicolumn{7}{c}{ErII}\\
   4023.00 &    0.21 &   $-$0.200 &   &   47.0   &   &    1.14 \\
   4043.59 &    0.32 &   $-$0.510 &   &   24.3   &   &    0.92 \\
   4051.14 &    0.38 &    0.090 &   &   39.5   &   &    0.80 \\
   4061.07 &    0.47 &    0.520 &   &   67.8   &   &    1.14 \\
   4109.44 &    0.00 &   $-$0.810 &   &   25.1   &   &    0.86 \\
   4113.83 &    0.18 &   $-$0.900 &   &   29.6   &   &    1.29 \\
   4133.35 &    0.32 &   $-$0.510 &   &   32.8   &   &    1.12 \\
   4156.07 &    0.18 &    0.180 &   &   69.9   &   &    1.17 \\
   4211.29 &    0.21 &   $-$0.720 &   &   30.0   &   &    1.15 \\
   4446.37 &    0.21 &   $-$0.590 &   &   46.5   &   &    1.38 \\
   4451.55 &    0.38 &    0.110 &   &   61.4   &   &    1.13 \\
   4451.98 &    0.00 &   $-$1.340 &   &   26.6   &   &    1.41 \\
   4567.60 &    0.21 &   $-$1.510 &   &    9.3   &   &    1.19 \\
   4579.32 &    0.74 &   $-$0.650 &   &   15.0   &   &    1.15 \\
   4645.77 &    0.56 &   $-$0.750 &   &   20.4   &   &    1.22 \\
   5092.78 &    0.38 &   $-$0.610 &   &   28.1   &   &    1.04 \\
   5192.62 &    1.14 &    0.310 &   &   35.2   &   &    1.10 \\
\enddata
\end{deluxetable}

\begin{deluxetable}{lrrcrrr}
\tablenum{1}
\tablewidth{0pt}
\tablecaption{Equivalent Widths}
\tablehead{
\colhead{$\lambda$} & \colhead {E.P.} & \colhead {log $gf$} & \colhead{Method} &\colhead {EW} & \colhead {HFS,} & \colhead {log $\epsilon$} \\
\colhead {\AA} & \colhead{(eV)}& \colhead{} & \colhead{} & \colhead {m\AA} & 
\colhead{IS?} & \colhead{} 
}
\startdata

   5200.12 &    0.56 &   $-$0.490 &   &   10.7   &   &    0.56 \\
   5212.35 &    0.21 &   $-$0.870 &   &   24.2   &   &    1.02 \\
\multicolumn{7}{c}{TmII}\\
   3462.20 &    0.00 &    0.030 & EW  &   32.5   &   &   $-$0.26 \\
   3700.26 &    0.03 &   $-$0.380 & SYN  &   23.7   &   &   $-$0.20 \\
   3701.36 &    0.00 &   $-$0.540 & SYN  &   16.6   &   &   $-$0.30 \\
   3761.91 &    0.00 &   $-$0.430 & SYN  &   17.2   &   &   $-$0.40 \\
\multicolumn{7}{c}{YbII}\\
   3694.20 &    0.00 &   $-$0.229 &  SYN &  118.6   & HFS,IS  &    0.76 \\
\multicolumn{7}{c}{LuII}\\
   3507.32 &    0.00 &   $-$1.540 & syn  &   51.4   &  HFS &   $-$0.10 \\
\multicolumn{7}{c}{HfII}\\
   3253.70 &    0.38 &   $-$0.580 & SYN   &   34.1   &   &    0.70 \\
   3389.83 &    0.45 &   $-$0.700 & SYN  &   26.7   &   &    0.62 \\
   3399.79 &    0.00 &   $-$0.490 & SYN  &   65.1   & HFS,IS  &    0.64 \\
   3505.22 &    1.04 &   $-$0.080 & SYN  &   30.2   &   &    0.70 \\
   3535.55 &    0.61 &   $-$0.540 & SYN  &   29.1   &   &    0.67 \\
   3569.03 &    0.79 &   $-$0.400 & SYN  &   27.7   &   &    0.67 \\
   3644.35 &    0.79 &   $-$0.480 & SYN  &   25.5   &   &    0.67 \\
   3719.28 &    0.61 &   $-$0.870 & SYN  &   26.5   &   &    0.82 \\
   3793.38 &    0.38 &   $-$0.950 & SYN  &   25.0   &   &    0.60 \\
   3918.09 &    0.45 &   $-$1.010 & SYN  &   23.4   &   &    0.67 \\
   4093.16 &    0.45 &   $-$1.090 & SYN  &   26.8   &   &    0.82 \\
\multicolumn{7}{c}{PbI}\\
   3683.47 &    0.97 &   $-$0.790 & EW &    42.8  & HFS,IS  &    2.43 \\
   4057.81 &    1.32 &   $-$0.481 & EW &    51.2  & HFS,IS  &    2.48 \\
\enddata
\tablenotetext{a}{All abundances given as [Element/Fe], except for Fe where
[Fe/H] is given.}
\end{deluxetable}